\documentclass{JHEP3}
\usepackage{amssymb,amsfonts,amsmath,amsthm,latexsym,graphicx,verbatim,epsfig}
\newcommand{\nn}{\nonumber}
\newcommand{\beq}{\begin{equation}}
\newcommand{\eeq}{\end{equation}}
\newcommand{\bea}{\begin{eqnarray}}
\newcommand{\eea}{\end{eqnarray}}

\newcommand\PSL{\mathop{\rm PSL}(2,\mathbb{Z})}
\author{Matteo Cardella  \\
    Racah Institute of Physics,
    Hebrew University of
    Jerusalem, 91904
    Israel.\\
  {\tt matteo@phys.huji.ac.il\\}
    }

\title{\center{A novel  method for computing torus amplitudes for  $\mathbb{Z}_{N}$ orbifolds without the unfolding  technique}}

\abstract{A novel method for computing  torus amplitudes in  orbifold compactifications is suggested.
 It applies  universally  for  every Abelian $\mathbb{Z}_{N}$ orbifold  without  requiring
  the unfolding technique.
  This method follows from the possibility of obtaining integrals over fundamental domains
  of every   Hecke congruence  subgroup $\Gamma_{0}[N]$
  by computing  contour integrals over one-dimensional curves uniformly  distributed in these domains.}

\keywords{Orbifolds, Congruence subgroups, Uniform distribution}

\begin{document}

\section{Introduction}

Superstring  orbifolds compactifications  are among the few examples where semi-realistic physics
emerges in  a complete string description.
By choosing an orbifold space  to compactify the superstring  which do not preserve any of its original supersymmetries,
 one can study   quantum effects induced by the infinite towers of string excitations.
This effects are encoded by the string free-energy given by the (worldsheet) torus amplitude.
This amplitude is  generically non zero after supersymmetry breaking, and  plays the role of a potential
in the geometric  string moduli.
Quantum  effects induced by supersymmetry breaking   include at times  generation of closed string tachyons,
and generically  uplifting of the string moduli.
The presence of closed string tachyons in  regions of the moduli space
 induce a break-down of the  analyticity of  the free-energy,
which signals   the presence of a phase transition involving  the space-time background itself. This
is a non-perturbative  process difficult to analyze except  under  very special circumstances.
The  non-tachyonic cases are   more under control,
although  afflicted by  the problem of moduli stabilization.
 Generically,  the torus  potential has runaway directions in the moduli space,  pushing  the system  towards its decompactification limits.
There are however examples where some or  all the geometric moduli can be stabilized in local minima of the torus potential \cite{Angelantonj:2006ut},\cite{Dine:2006gx},\cite{Angelantonj:2008fz}.

\vspace{.3 cm}

\begin{figure}
\centering
\includegraphics[width=3cm]{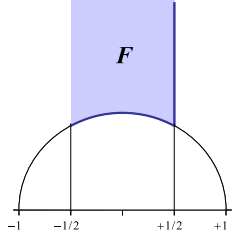}
\caption[]{The standard fundamental region  $\mathcal{F} = \{|\tau|> 1,  -1/2 \le \tau_1 < 1/2 \} $ of the torus modular group
$\Gamma \sim  \PSL$ in the upper complex plane $\mathbb{H}$.}
\label{domain}
\end{figure}

In order to obtain the potential in the string moduli one has to   compute the torus amplitude.
This is given by  an integral in the complex worldsheet  torus parameter $\tau$
 over a fundamental region $\mathcal{F}$, (shown in figure \ref{domain}) of the torus modular group $\Gamma \sim \PSL$.

     For a $\mathbb{Z}_{p}$-orbifold with $p$ prime integer the torus amplitude has the following structure\footnote{In $\mathbb{Z}_{N}$-orbifolds  with non-prime $N$ the structure of the torus potential  is slightly
      more complicate, due to terms invariant under the $\mathbb{Z}_{N}$  cyclic subgroups. We will discuss the general form for every $N$ later on.}

\bea
V_{p} &\sim& -  \int_{\mathcal{F}}\frac{d^{2}\tau}{\tau_{2}^{2}}\left(1 + S + TS + ... + T^{p-1}S     \right) \nn \\ && \cdot \frac{1}{\tau_{2}^{-1 + d/2}} Str_{(\mathcal{H}_{L}\times\mathcal{H}_{R})} \left(\frac{1/p +g+...+ g^{p-1}}{p} q^{L_0 - c/24} \bar{q}^{\bar{L}_{0} - \bar{c}/24 } \right)  \nn \\
&=& \int_{\mathcal{F}}\frac{d^{2}\tau}{\tau_{2}^{2}}\left(1 + S + TS + ... + T^{p-1}S \right)\left(\frac{f_{0,0}}{p} + f_{1,0} + ... + f_{p-1,0}  \right) \nn \\
\label{potential}
\eea

where   $q=e^{2i\pi \tau}$, and $d$ is the number of non-compact  dimensions. $g$ is the orbifold operator with a definite action on the superstring states belonging to $\mathcal{H}_{L}\times\mathcal{H}_{R}$.
   $g$ generates the  $\mathbb{Z}_{p}$ cyclic group      ($g^{p} = 1$),  $\mathbb{Z}_p \sim \{ \{0,1,...,p-1 \},  + (mod \  p) \}$.

\vspace{.2 cm}
In (\ref{potential}) we have used the following notation

\beq
f_{i,j}(\tau, \bar{\tau}) = \frac{1}{p\tau_{2}^{1-d/2}}Str_{(\mathcal{H}_{L}\times\mathcal{H}_{R})_{j}}\left(g^{i} q^{L_0 - c/24} \bar{q}^{\bar{L}_{0} - \bar{c}/24 } \right), \label{fi}
\eeq
for the contributions from the $j$-twisted strings
 with the $g^{i}$ insertion in the supertrace.
 All the worldsheet fields in the $j$-twisted sector satisfy the boundary conditions
 \beq
 \phi(\sigma + 2\pi, t) = g^{j}\phi(\sigma , t).
 \eeq

  The $g^{i}$ insertion in eq. (\ref{fi})  produces a  twisting in the
fields boundary condition along  the $t$-homology cycle of the worldsheet torus.

 The modular transformations in (\ref{potential}), where $T: \tau \rightarrow \tau + 1$ and $S: \tau \rightarrow -1/\tau$,
are required to produce new terms which  complete  a modular invariant multiplet\footnote{$f_{0,0}(\tau, \bar{\tau})$ describes the closed string theory partition function before the orbifold  compactification. If the original string theory is supersymmetric then this term  is identically zero.}.
The action of the generator $S$ and $T$ on the functions $f_{i,j}(\tau, \bar{\tau})$
is given by (mod($p$))
\bea
S: f_{i,j}(\tau, \bar{\tau}) &\rightarrow&  f_{-j,i}(\tau, \bar{\tau})  \nn \\
T: f_{i,j}(\tau, \bar{\tau}) &\rightarrow&  f_{i+j,j}(\tau, \bar{\tau}).  \nn
\eea

\vspace{.2 cm}

 One can check that  the  set of  $p^2 - 1$ terms  in the integral (\ref{potential})
 form a modular invariant multiplet. In fact,  the set of transformations $(S,TS,...,T^{p-1}S)$ by acting on  each of the
 $T$-invariant   $p-1$ terms  $f_{i,0}$,
 generate a $(p - 1)$-dimensional $T$-invariant multiplet.  Terms in distinct  multiplets are then connected
by  $S$ transformations.

 \vspace{.2 cm}
 By performing a change of   integration variable in (\ref{potential}) one can rewrite the torus amplitude $V_{p}$ as follows
\beq
V_{p} \sim -  \int_{\mathcal{F}_{p}}\frac{d^{2}\tau}{\tau_{2}^{2}}\left(\frac{f_{0,0}}{p} + f_{1,0} + ... + f_{p-1,0}    \right). \label{potential2}
\eeq

where $\mathcal{F}_{p} = \mathcal{F} \cup_{i=1}^{p-1}ST^{i}(\mathcal{F})$. This new integration region is a fundamental domain for the congruence subgroup $\Gamma_{0}[p] \subset \Gamma$, given by $\PSL$ matrices  of the form

\beq
\begin{pmatrix}
 a  &  b \\
 pc  &  pd+k  \\
\end{pmatrix},
\eeq

with $(pc,pd+k) := MCD(pc,pd+k)  = 1$.

\vspace{.4 cm}

 In the literature  computation of the $\tau$ integral over the region $\mathcal{F}_{p}$ in (\ref{potential2}) is usually carried
 on by using the unfolding technique \cite{McClain},\cite{fBrien},\cite{Itoyama},\cite{Dixon},\cite{MS},\cite{GNS},\cite{Tra},\cite{KKPR}.
  In toroidal orbifolds  for a compactification down to $d$-dimensions,
 lattice  states given by the  $d$-dimensional momentum quantum number $\vec{m}$
  and  the $d$-dimensional winding number $\vec{n}$ are used to unfold the $\mathcal{F}_{p}$ domain.
   These quantum numbers   can be arranged to form a representation of  a  subgroup $G$ of  $GL(10 - d,\mathbb{Z})$.
   By computing the orbits of $\Gamma_{0}[p]$ in $G$,
   the original integral (\ref{potential2}) on the domain $\mathcal{F}_{p}$
    can be reduced to an integral  over the strip $\mathcal{S} = [-1/2,1/2] \times [0,\infty)$
involving  as many terms as the number of independent  orbits of  $\Gamma_{0}[p]$ in $G$. For a generic $\mathbb{Z}_{N}$
 this method can be quite  complicate to follow,  and the tricks to be used to obtain the final  unfolded integral
 depend on the dimension of the subgroup  $G \subset GL(10 - d,\mathbb{Z})$ \cite{Dixon},\cite{KKPR}.
The general  method for unfolding the integration domain for a generic $\mathbb{Z}_{N}$ orbifold
is studied  in \cite{Tra}.

\vspace{.4 cm}
Here we propose  a different way for computing integrals over fundamental regions $\mathcal{F}_{p}$ of the congruence
subgroups $\Gamma_{0}[p]$ of the kind of (\ref{potential2}). Instead of unfolding
$\mathcal{F}_{p}$ into the strip $\mathcal{S}$, we trade the integral over $\mathcal{F}_{p}$ for a
contour   integrals over a (one-dimensional) curve which is uniformly  distributed in $\mathcal{F}_{p}$.
 Uniform distributions property of one-dimensional curves in homogenous space with
negative curvature  has been extensively
studied in the mathematics literature \cite{F},\cite{Dani},\cite{Ratner1},\cite{Ratner2} and quite general theorems have been obtained.

  In  appendix we give our   proof of a uniform distribution theorem
  for  $\mathbb{H}/ \Gamma_{0}[N]$  hyperbolic spaces based on elementary function analysis.
   This theorem  states that for every congruence subgroup $\Gamma_{0}[N]$ with fundamental
 region $\mathcal{F}_{N}$ in the upper complex plane $\mathbb{H}$, there is  a (one-dimensional) curve
 which is dense and uniformly distributed in $\mathcal{F}_{N}$.
This curve appears  is the image in $\mathcal{F}_{N}$
 of the infinite radius horocycle\footnote{A horocycle in the upper  hyperbolic plane $\mathbb{H}$ is a circle
 tangent to the real axis. In the infinite radius limit a horocycle degenerates into the real axis.}  in the upper hyperbolic plane $\mathbb{H}$.

A sequence of horocycles $\{h_n \}_{n\in \mathbb{N}}$ converging to the infinite
radius horocycle, (the real axis),
have their image  curves $\{\gamma_n \}_{n\in \mathbb{N}}$ in  $\mathcal{F}_{N}$
which tend to become uniform distributed in $\mathcal{F}_{N}$ for $n \rightarrow \infty$.
Therefore\footnote{Equation (\ref{int}) shows that the horocycle flow is \emph{ergotic} on the hyperbolic space $\mathbb{H}/\Gamma_{0}[N]$.} for enough regular function  $f(\tau, \bar{\tau})$

\beq
\lim_{n\rightarrow \infty} \frac{1}{L(\gamma_n)} \oint_{\gamma_n}ds f(\tau, \bar{\tau})=
\frac{1}{A(\mathcal{F}_{N})}\int_{\mathcal{F}_{N}}\frac{d^{2}\tau}{\tau_{2}^2}f(\tau, \bar{\tau}), \label{int}
\eeq

where $L(\gamma_n)$ is the length of $\gamma_n$ computed by the hyperbolic metric

\beq
L(\gamma_n) = \oint_{\gamma_n}ds =     \oint_{\gamma_n}\frac{\sqrt{d\tau_{1}^2 + d\tau_{1}^2}}{\tau_2}.
\eeq

In eq. (\ref{int}) the integral over a fundamental region $\mathcal{F}_N$
  is normalized  by the   area $A(\mathcal{F}_{N})$ of the hyperbolic polygon $\mathcal{F}_{N}$.

\beq
A(\mathcal{F}_{N}) = \int_{\mathcal{F}_N}\frac{d\tau_1 d\tau_2}{\tau_{2}^2}.
\eeq

\vspace{.2 cm}

Since the limiting curve $\gamma_{\infty}$ in (\ref{int})
is the image of the infinite radius horocycle (the real axis) then for every enough regular
 $\Gamma_{0}[N]$-invariant function $f$ \footnote{ The regularity conditions on the function $f$
 are given  in appendix.  The same relation for functions invariant under the full modular group $\Gamma \sim \PSL$
integrated over a fundamental $\mathcal{F}$ has been used in \cite{Kutasov:1990sv} to study the asymptotic cancelation
among bosonic and fermionic closed string excitations in non-tachyonic backgrounds, (see also \cite{Kutasov:1991pv},\cite{Dienes:1994es},\cite{Dienes:1995pm}).}

\beq
\int_{\mathcal{F}_{N}}\frac{d^{2}\tau}{\tau_{2}^2}f(\tau, \bar{\tau}) = A(\mathcal{F}_{N})\lim_{\tau_{2} \rightarrow 0} \int_{-1/2}^{1/2} d \tau_{1}  f(\tau, \bar{\tau}). \label{th}
\eeq

\vspace{.6 cm}

This result provides  an alternative way  for computing the torus amplitude (\ref{potential2}),
and more generally the torus amplitude  for every  $\mathbb{Z}_{N}$ orbifold, $N \in \mathbb{N}$.

\vspace{.2 cm}

The organization of the rest of the paper is the following:
in the next section we start by considering   specific examples such as the
$\mathbb{Z}_{4}$ and  $\mathbb{Z}_{6}$ orbifolds and  illustrate in details the construction
of the modular invariant multiplets. Then we show the equivalence of the modular integral for the torus amplitude
to a $\tau_{2} \rightarrow  0$ limit  of the untwisted sector partition functions, modified by
coefficients  depending  on the dimensions of the cyclic subgroups of $\mathbb{Z}_{4}$ and $\mathbb{Z}_{6}$.
  We then  provide  the general formula for the torus amplitude, valid for a generic $\mathbb{Z}_{N}$.
 The proof of the uniform distribution theorem  is given in the appendix.

\vspace{.1 cm}

\section{The torus amplitude for a  generic $\mathbb{Z}_{N}$ orbifold}

\subsection{The  $\mathbb{Z}_{4}$ case}

\vspace{.2 cm}

For a $\mathbb{Z}_{4}$ orbifold  the torus amplitude has the following structure\footnote{In the following we omit the contribution to the free-energy
from the uncompactified theory $f_{0,0}$. This contribution is zero if the original theory is supersymmetric. Otherwise in all the following formulae
the extra term $\frac{1}{N}\int_{\mathcal{F}}\frac{d^{2}\tau}{\tau_{2}^2}f_{0,0}$ has to be added, where $N$ is the order of the $\mathbb{Z}_N$ orbifold. }

\beq
V_4 \sim \int_{\mathcal{F}}\frac{d^{2}\tau}{\tau_{2}^2} (1 + S + TS )(1+ ST^{2}S )\left(f_{1,0} + \frac{1}{2}f_{2,0} + f_{3,0}\right). \label{VZ4}
\eeq
 $f_{1,0}$  and $f_{3,0}$  are $\Gamma_{0}[4]$ invariant, while
 $f_{2,0}$  is invariant under the larger congruence subgroup $\Gamma_{0}[2] \supset \Gamma_{0}[4]$.

Since  $(1 + S + TS + T^2 S + T^3 S)f_{2,0} = 2(f_{2,0} + f_{0,2} + f_{2,2})$, the terms  $f_{1,2}$ and $f_{3,2}$
 are obtained from $f_{1,0}$  and $f_{3,0}$ through a $ST^{2}S$ transformation.

\vspace{.4 cm}

By a change of integration variable,  eq. (\ref{VZ4}) can be reduced to
\beq
V_4 \sim \int_{\mathcal{F}_{4}}\frac{d^{2}\tau}{\tau_{2}^2}\left(f_{1,0} + \frac{1}{2}f_{2,0} + f_{3,0}\right), \label{VZ4bis}
\eeq

where $\mathcal{F}_{4}$ is a fundamental domain for $\Gamma_{0}[4]$

\beq
\mathcal{F}_{4} = \mathcal{F} \cup \bigcup_{i=0}^{3}ST^{i}(\mathcal{F}) \bigcup ST^{2}S(\mathcal{F}). \label{F4}
\eeq

By using the uniform distribution of the infinite radius horocycle in $\mathcal{F}_4$ one can then
express the torus  potential  for a generic  $\mathbb{Z}_{4}$ orbifold (\ref{VZ4bis})  as the following $\tau_{2} \rightarrow 0$ limit

\beq
V_4 \sim  6\cdot \frac{\pi}{3} \lim_{\tau_{2} \rightarrow 0} \int_{-1/2}^{1/2}d\tau_{1}\left(f_{1,0} + \frac{1}{2}f_{2,0} + f_{3,0}\right)(\tau, \bar{\tau}), \label{VZ4limit}
\eeq

 where  the factor $6\pi/3$ is equal to the invariant area of the fundamental region $\mathcal{F}_{4}$ of $\Gamma_{0}[4]$\footnote{The invariant area of a fundamental domain  $\mathcal{F}$ of $\Gamma$ is $\pi/3$, and it can be obtained by recalling   that 
  the area $A$ of an hyperbolic triangle is  given by $A = \pi - \sum_{i=1}^{3}\alpha_i$, where $\alpha_i$ are its internal angles.
   $\mathcal{F}_4$ is covered by six  fundamental regions of $\Gamma$ as  shown in eq. (\ref{F4}).}. Notice in eq. (\ref{VZ4limit}) the presence of the factor $1/2$ in front of $f_{2,0}$. This is connected
 with the invariance of this term under the larger congruence subgroup $\Gamma_{0}[2]$.   In the general formula to be given below
 for a $\mathbb{Z}_N$ orbifold when $N$ is not prime, rational coefficient will appear in front of terms which are
 invariant under the cyclic subgroups of $\mathbb{Z}_N$. Before writing the general formula we study in the next section
 the $\mathbb{Z}_6$ example.

\subsection{The  $\mathbb{Z}_{6}$  case}

The torus amplitude is given by
\bea
V_6 &\sim&  \int_{\mathcal{F}}\frac{d^{2}\tau}{\tau_{2}^2}\Bigg[\left(1 + \sum_{i=0}^{5}T^{i}S  + (1 + S + TS)T^{3}S + (1+S+TS +T^{2}S)T^{4}S \right)
(f_{1,0} + f_{5,0})  \nn \\  &+&
(1 + S + T S + T^{2}S )(f_{2,0} + f_{4,0}) +  (1 + S + T S  )f_{3,0} \Bigg].
\eea

The above structure follows from  the  $\Gamma_{0}[3]$ invariance  of $(f_{2,0} + f_{4,0})$,
and the $\Gamma_{0}[2]$ invariance  of $f_{3,0}$.

\vspace{.2 cm}

The amplitude can  be rewritten as

\bea
V_6 \sim \int_{\mathcal{F}_{6}}\frac{d^{2}\tau}{\tau_{2}^2}(f_{1,0} + f_{5,0}) + \int_{\mathcal{F}_{3}}\frac{d^{2}\tau}{\tau_{2}^2}(f_{2,0} + f_{4,0})
+ \int_{\mathcal{F}_{2}}\frac{d^{2}\tau}{\tau_{2}^2}f_{3,0}. \label{VZ6}
\eea

By using uniform distribution property one can express the same amplitude as the following limit

\beq
V_6  \sim 12 \cdot \frac{\pi}{3}  \lim_{\tau_{2} \rightarrow 0} \int_{-1/2}^{1/2}d\tau_{1} \left( f_{1,0} + \frac{1}{3}f_{2,0} +
 \frac{1}{4}f_{3,0} + \frac{1}{3}f_{4,0} + f_{5,0} \right), \label{VZ6limit}
\eeq

where $12 \pi/3$ is the hyperbolic area of a fundamental region of $\Gamma_{0}[6]$.
 \vspace{.4 cm}

 \subsection{Amplitude for a generic $\mathbb{Z}_{N}$ orbifold}

The  analysis  in the previous section for the $\mathbb{Z}_4$ and  $\mathbb{Z}_6$ orbifolds
suggests the way  for obtaining  a decomposition of  a generic $\mathbb{Z}_N$  torus amplitude as a sum of integrals over
 fundamental regions of congruence subgroups $\Gamma_{0}[q]$, $2 \le q \le N$.
This decomposition together  with the theorem on uniform distribution\footnote{The theorem in the appendix gives a finite correction term
 $24\frac{\pi}{3}\sum_{i=1}^{n_{c}(N)}c_{i}\beta_{i}$ in eq. \ref{KS2thesis} in the presence of divergences at the cusps of $\mathcal{F}_{N}$.
  This divergences correspond to untwisted and twisted unphysical tachyons in the orbifold partition function,
  i.e. states that are eliminated  by level matching through $\tau_1$ integration of the partition function.
    If one compactifies  all the space-time dimensions except $2$, ($d=2$) one recovers this finite correction.
  Therefore in $d=2$ in the torus amplitude this correction appears multiplied by $1/VOL(8)$, where $VOL(8)$ is the volume of the eight-dimensional compact space.
  This shows that  this finite correction vanishes  in every orbifold compactification with $d>2$.}
gives  the following formula  for the  torus amplitude in
a non-tachyonic $\mathbb{Z}_{N}$ orbifold

\beq
V_{N} \sim   \lim_{\tau_{2}\rightarrow 0} \int_{-1/2}^{1/2}d\tau_{1} \sum_{l=1}^{N-1}A\left(\mathcal{F}_{n(l)}\right)f_{l,0}(\tau, \bar{\tau}),\label{general}
\eeq
where the integer numbers $0< n(l) <N$ are solutions of the following equation\footnote{$n(l)$ is the dimension of the cyclic subgroup of $\mathbb{Z}_{N}$ generated by the element $g^{l}$, when $(l,N)>1$.}

\beq
l \cdot  n(l) = N, \ \ mod(N),  \label{mod}
\eeq
and $A(\mathcal{F}_{r})$ is the area of the fundamental region $\mathcal{F}_{r}$ of the congruence subgroup $\Gamma_{0}[r]$\footnote{See the appendix for a derivation of the  formula for the area  $A(\mathcal{F}_{r})$ of the fundamental region $\mathcal{F}_r$.} which can be computed by
\beq
A(\mathcal{F}_{r}) =  \frac{2r^2}{\pi \varphi(r)}\sum_{(k,r)=1}\sum_{n=0}^{\infty}\frac{1}{(r n + k)^{2}}.
\eeq
In the last formula $\varphi(r)$  is the Euler totient phi-function, which  counts the number of integers  $k$, $1\le k <r$   coprime with $r$,
$(k,r) =1$.
Given the $r$ decomposition in prime factors $r = p_{1}^{s_1}\cdot ... \cdot p_{q}^{s_q}$, $\varphi(r)$ can be computed by the following
Euler product
\beq
\varphi(r) = \prod_{p|r}\left(1 - \frac{1}{p}   \right),
\eeq
where $p|r$ indicates that $p$ is a divisor of $r$.

\vspace{.1 cm}
In equation (\ref{mod}) if $l$ is coprime with $N$, $(l,N) = 1$, then $n(l)=N$ and $g^{l}$ generates the full $\mathbb{Z}_N$. If $(l,N) > 1$ then $n(l)$ is the common factor
between $l$ and $N$, $n(l)<N$, and $g^{l}$ generates the cyclic subgroup $\mathbb{Z}_{n(l)} \subset \mathbb{Z}_N$.
 In this last case  the untwisted terms $f_{l,0}(\tau, \bar{\tau})$, are  invariant under the congruence subgroup $\Gamma_{0}[n(l)]$,  with fundamental domain $\mathcal{F}_{n(l)}$.
 This was the case for the unwisted terms which appeared dressed by fractional coefficients
for the   $\mathbb{Z}_{4}$ orbifold  in eq. (\ref{VZ4limit}) and for  the $\mathbb{Z}_{6}$ orbifold in eq. (\ref{VZ6limit}).
This fractional coefficients are the ratios $A(\mathcal{F}_{n(l)})/A(\mathcal{F}_{N})$
of the areas of the fundamental regions of $\Gamma_{0}[n(l)]$ and $\Gamma_{0}[N]$, which are rational numbers since
 every congruence subgroup is covered by a finite number of fundamental regions of $\Gamma$.

\acknowledgments
The Author is grateful to Carlo Angelantonj for useful comments and reading of the manuscript.
The Author   thanks Ehud de Shalit, Hershel Farkas, Erez Lapid,  Elon Lindenstrauss and Ron Livne
 for their precious  expertise   on modular functions.
This work  is  partially  supported by Superstring  Marie Curie Training Network under the
contract MRTN-CT-2004-512194.

\vspace{.5 cm}

\appendix

\section{Uniform distribution of curves in the fundamental regions of the Hecke congruence subgroups}

For every integer $N > 1$, the congruence subgroup  $\Gamma_{0}[N] \subset \Gamma$ is represented by matrices in  $\PSL$  with $c=0$, (mod $N$).
These  matrices have the form
 \beq
  \begin{pmatrix}
 a  &  b \\
 Nc  &  Nd + k  \\
\end{pmatrix},
\eeq

with $(Nc,Nd + k) = 1$.

\begin{figure}
\centering
\includegraphics[height=3cm]{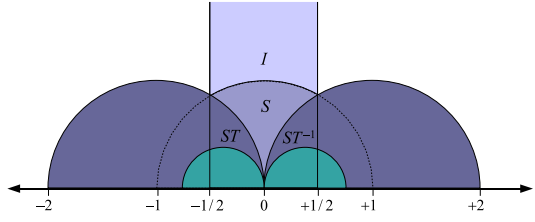}
\caption[]{The region $I \cup S \cup ST$ is a fundamental domain for $\Gamma_{0}[2]$ }
\label{domainbis}
\end{figure}

A Fundamental region of $\Gamma_{0}[N]$ on the hyperbolic upper plane $\mathcal{F}_N = \mathbb{H}/ \Gamma_{0}[N]$
is given by the union of a fundamental region $\mathcal{F}$ of the full modular group $\Gamma$
with the images of $\mathcal{F}$ through all the transformations in $(\Gamma - \Gamma_{0}[N])/T$.
$\mathcal{F}_N$ is an hyperbolic polygon whose vertexes in $z=i\infty$ and in points in $\mathbb{Q} \cup [-1/2,1/2]$
are called cusps, (the fundamental region $\mathcal{F}_2$ of $\Gamma_{0}[2]$ is shown in figure \ref{domainbis}.)

\vspace{.2 cm}

Here we prove that for every congruence subgroup $\Gamma_{0}[N] \subset \Gamma$, the image of the infinite radius horocycle \footnote{A horocycle is
a circle tangent to the real axis and contained in $\mathbb{H}$.
Every horocycle  $\mathcal{H}$ of radius $R$ has an image curve $\gamma_{R}$  under $\Gamma_{0}[N]$
 transformation fully contained in $\mathcal{F}_N = \mathbb{H}/ \Gamma_{0}[N]$. In the infinite radius limit $R \rightarrow \infty$
 every horocycle degenerates to the real axis.} through $\Gamma_{0}[N]$
transformations is uniformly distributed in the fundamental region $\mathcal{F}_N = \mathbb{H}/ \Gamma_{0}[N]$.
To this purpose we will show that for every regular enough\footnote{Respecting the conditions of the theorem displayed below.}  $\Gamma_{0}[N]$ invariant function
$f(\tau, \bar{\tau})$

\beq
\frac{1}{L(\gamma_{\infty})}\oint_{\gamma_{\infty}^{N}}ds f  =                 \lim_{\tau_2 \rightarrow 0}\int_{-1/2}^{1/2}d\tau_{1} f  \rightarrow \frac{1}{A(\mathcal{F}_n)}\int_{\mathcal{F}_N} \frac{d\tau_1 d\tau_2}{\tau_{2}^2}f, \label{unif}
\eeq

where the  upper plane $\mathbb{H}$  hyperbolic metric is given by
\beq
ds^2 = \frac{d\tau_{1}^2 + d\tau_{2}^2}{\tau_{2}^2}.
\eeq

  In eq. (\ref{unif})  $\gamma_{\infty}^{N} \subset \mathcal{F}_N$ denotes the image curve of the infinite radius horocycle,
  and $L(\gamma)$ denotes  the hyperbolic length of a  curve $\gamma$
\beq
L(\gamma) = \oint_{\gamma}ds = \oint_{\gamma} \frac{\sqrt{d\tau_{1}^2 + d\tau_{2}^2}}{\tau_2}.
\eeq

\vspace{.2 cm}

\begin{itemize}
\item i) Let $f(\tau,\bar{\tau})$ be a  function invariant under $\Gamma_{0}[N]$,  finite over the fundamental domain
$\mathcal{F}_{N}$ of $\Gamma_{0}[N]$, except possibly at the cusps of $\mathcal{F}_{N}$, which include $\tau = i\infty$ and  points
in $\mathbb{Q} \cap [-1/2,1/2]$\footnote{$\mathcal{F}_N$ has cusps in $\mathbb{Q} \cap [-1/2,1/2]$ which are the images of the point $\tau = \infty$ through a finite number of modular transformations $\{\mathcal{M}_{i} \}_{i\le I}$ with the following property. For
 every $\mathcal{M} \in \Gamma$, $\mathcal{M} = \mathcal{M}_{N}\mathcal{M}_{i}$ for $\mathcal{M}_{N} \in \Gamma_{0}[N]$ and $\mathcal{M}_{i}$ in the list  $\{\mathcal{M}_{i} \}_{i\le I}$. When $N=p$ is prime $\{\mathcal{M}_{i} \}_{i\le I} = \{ST^{i}  \}_{1 \le i \le p-1}$
 and the only cusps of $\mathcal{F}_p$ on the real axis is in $\tau = 0$, since $ST^{i}(\infty) = 0$. When $N$ is non-prime
 $\mathcal{F}_N$ has extra cusps on the real axis in non-vanishing rational points inside $[-1/2,1/2]$.}.

\item ii) Let  the integral on $\mathcal{F}_{N}$ of $f(\tau,\bar{\tau})$  be convergent
\beq
\left| \int_{\mathcal{F}_{N}}\frac{d^{2}\tau}{\tau_{2}^2}f(\tau,\bar{\tau}) \right| < \infty. \label{int2}
\eeq

\item iii)
$f$ has the following Fourier expansion
\beq
f(q, \bar{q}) = \sum_{ i =1}^{n_{c}(N)} \frac{c_i}{q - e^{2\pi i \tau_{i}}}      + regulars,
\eeq
where $q=e^{2\pi i \tau}$, $n_{c}(N)$ is the number of cusps of $\mathcal{F}_{N}$, and $\tau_i$ the locations
of the cusps.
\end{itemize}

\vspace{.1 cm}

 Then:

\beq
 \int_{\mathcal{F}_{N}}\frac{d^{2}\tau}{\tau_{2}^2}f(\tau,\bar{\tau}) = A(\mathcal{F}_{N}) \lim_{\tau_2 \rightarrow 0}\int_{-1/2 }^{1/2 }d\tau_{1}f(\tau,\bar{\tau}) + 24\frac{\pi}{3}\sum_{i=1}^{n_{c}(N)}c_{i}\beta_{i},\label{KS2thesis}
\eeq

 where in the above equation
\beq
A(\mathcal{F}_{N}) =  \frac{2N^2}{\pi \varphi(N)}\sum_{(k,N)=1}\sum_{n=0}^{\infty}\frac{1}{( nN + k)^{2}}. \label{area}
\eeq
  is the modular invariant area of the fundamental  region $\mathcal{F}_{N}$ of $\Gamma_{0}[N]$\footnote{$A(\mathcal{F}_{N})$ is of the form
$\mathcal{N}(N)\pi/3$ where $\mathcal{N}(N)$ is an integer. This follows from the fact that $\mathcal{F}_{N}$ can be covered by finite number  of  fundamental domains of the full modular group $\Gamma$ with invariant area $A(\mathcal{F})= \pi/3$. Each tile corresponds to the image of $\mathcal{F}$ through
modular transformations in the set $\Gamma - \Gamma_{0}[N]$. For $N$ prime there are $N + 1$ tiles, $c(p) = p+1$ for prime $p$.},
and $\beta_i$ is the number of fundamental regions of the full modular group $\Gamma$ in the tassellation of $\mathcal{F}_{N}$
which have the same cusp $\tau_i$ \footnote{For example in $\mathcal{F}_{2}$:   $\beta(\tau =  \infty) = 1$, and $\beta(\tau = 0) = 2$,
 as shown in figure \ref{domainbis}.}. In eq. (\ref{area})  $\varphi(N)$  is the Euler totient phi-function\footnote{ $\varphi(r) $    counts how many numbers $k$, $1\le k <N$  are coprime with $N$,
$(k,N) =1$.
Given the  decomposition in prime factors $N = p_{1}^{s_1}\cdot ... \cdot p_{l}^{s_l}$, $\varphi(N)$ can be computed by the following
Euler product
$$\varphi(N) = \prod_{p|N}\left(1 - \frac{1}{p}   \right),$$ where $p|N$ indicates that $p$ is a divisor of $N$.}.

\proof
\vspace{1 cm}

We consider the following $\Gamma_{0}[N]$-invariant  auxiliary function
\beq
h_{N}(\tau, R) = \sum_{(k,N)=1}^{N-1}\sum_{m=-\infty}^{\infty}\sum_{n=-\infty}^{\infty} e^{-\frac{\pi R^{2}}{N^{2}\tau_2}|nN\tau + m N + k|^2}.
\eeq

By using Poisson resummation formula one can prove the following

\beq
\sum_{m=-\infty}^{\infty}\sum_{n=-\infty}^{\infty}e^{2\pi i m k/N }e^{-\frac{\pi }{R^{2}\tau_2}|n\tau + m|^2} = R^{2}\sum_{m=-\infty}^{\infty}\sum_{n=-\infty}^{\infty} e^{-\frac{\pi R^{2}}{N^{2}\tau_2}|nN\tau + m N  + k|^2}.
\eeq

From the previous two  relation by taking the limit $R \rightarrow  0$ one obtains the following identity

\beq
\frac{1}{\varphi(N)}\lim_{R \rightarrow 0}R^{2}\sum_{(k,N)=1}^{N-1} \sum_{m=-\infty}^{\infty}\sum_{n=-\infty}^{\infty} e^{-\frac{\pi R^{2}}{p^{2}\tau_2}|nN\tau + m N +k|^2}= 1.
\eeq

\vspace{.2 cm}

By using  the previous identity one therefore  has

\beq
\int_{\mathcal{F}_{N}}\frac{d^{2}\tau}{\tau_{2}^2}f(\tau,\bar{\tau}) = \frac{1}{\varphi(N)}\lim_{R \rightarrow 0}R^{2} \int_{\mathcal{F}_{N}}\frac{d^{2}\tau}{\tau_{2}^2}f(\tau,\bar{\tau})\sum_{m=-\infty}^{\infty}\sum_{n=-\infty}^{\infty}\sum_{(k,N)=1}^{N-1}e^{-\frac{\pi R^{2}}{N^{2}\tau_2}|nN\tau + m N +k|^2}. \label{a}
\eeq

Let us decompose $n N = (Nr + s)Nc$ and $m N + k = (Nr + s)(Nd + k')$ where $Nr + s = (nN, mn + k )$
  and therefore (N,s)=1 with $1 \le s \le N - 1$.  $Nc$ and $Nd + k'$ are therefore  coprime  $(Nc, Nd + k')=1$.

 With the above decomposition   (\ref{a}) becomes

\beq
\int_{\mathcal{F}_{N}}\frac{d^{2}\tau}{\tau_{2}^2}f(\tau,\bar{\tau}) = \frac{1}{\varphi(N)}\lim_{R \rightarrow 0}R^{2} \int_{\mathcal{F}_{N}}\frac{d^{2}\tau}{\tau_{2}^2}f(\tau,\bar{\tau}) \sum_{(s,N)=1}^{N-1} \sum_{r=-\infty}^{\infty}\sum_{c,d \in \mathbb{Z}} \sum_{(k,N)=1}^{N-1}e^{-\pi R^{2}(Nr + s)^2 \frac{|Nc \tau +  Nd  + k|^2}{N^{2}\tau_2}}. \label{a2}
\eeq

Notice at the exponent the images of $\tau_2$ under  $\Gamma_{0}[N]$ transformations. In fact,
under a generic $\Gamma_{0}[N]$ transformation  given by a matrix with lower row $(Nc, \ \   Nd + k )$, $\tau_2$ is mapped to

\beq
 \tau_2 \rightarrow \frac{\tau_2}{|Nc \tau +  Nd  + k|^2}.
\eeq

\vspace{.2 cm}
Moreover, since left multiplication by $T^{q}$, $q\in \mathbb{Z}$ of a generic $\Gamma_{0}[N]$ matrix  leaves its lower row invariant
\beq
T^{q} \begin{pmatrix}
 a  &  b \\
 Nc  &  Nd+k  \\
\end{pmatrix}
 = \begin{pmatrix}
 1  &  q \\
 0  &  1  \\
\end{pmatrix}
\begin{pmatrix}
 a  &  b \\
 Nc  &  Nd+k  \\
\end{pmatrix}
=
\begin{pmatrix}
 a'  &  b' \\
 Nc  &  Nd+k  \\
\end{pmatrix}, \label{T}
\eeq

 the set of $\Gamma_{0}[N]$ matrixes at the exponent in (\ref{a2}) form twice\footnote{Twice, since in (\ref{T}) there is an identification $k \sim N - k$
  which follows from the fact that the modular group and  all its congruence subgroups are projective.} a representation of $\Gamma_{0}[N]/T$.

\vspace{.3 cm}

For a given $c$,$d$ and $k$ we call $M_{c,d,k}$
\beq
M_{c,d,k} =
\begin{pmatrix}
 a  &  b \\
 Nc  &  Nd+k  \\
\end{pmatrix},
\eeq
the  matrix which maps $\tau \in \mathcal{F}_{N}$ into a point
  $M_{c,d,k}\tau \in \mathcal{S} - \mathcal{F}_N$, where      $\mathcal{S} = [-1/2, 1/2]\times [0, \infty)$.

\vspace{.2 cm}
 By changing integration variable $\tau \rightarrow M_{c,d,k}^{-1}\tau$  in the generic term in the r.h.s.
of (\ref{a}) one finds

 \beq
\int_{\mathcal{F}_{N}}\frac{d^2 \tau}{\tau_{2}^{2}}f(\tau,\bar{\tau})e^{-\frac{\pi R^2(Nr + s)^{2}}{N^{2}\tau_2}| Nc \tau +  Nd + k|^2}
 = \int_{M_{c,d,k}(\mathcal{F}_{N})}
 \frac{d^2 \tau}{\tau_{2}^{2}}f(\tau,\bar{\tau})e^{-\frac{\pi (Nr + s)^{2} R^2}{N^{2}\tau_2}}. \label{cvar}
\eeq

The union of all the $\{ M_{c,d,k} \}$ span twice the coset $\Gamma_{0}[N]/T$,
and therefore
\beq
\mathcal{F}_{N} \cup \bigcup_{c,d,k}M_{c,d,k}(\mathcal{F}_{N})
\eeq
is a double tassellation of the strip $\mathcal{S} = [-1/2,1/2]\times [0,\infty)$, whose tiles are an infinite set of fundamental regions of $\Gamma_{0}[N]$.

\vspace{.2 cm}

Thus by changing integration variable $\tau \rightarrow M_{c,d,k}^{-1}\tau$
term by term in the r.h.s. of eq. (\ref{a2}), one should recover

\bea
 \int_{\mathcal{F}_{N}}\frac{d^2 \tau}{\tau_{2}^{2}}f(\tau,\bar{\tau})
 %&=& \frac{1}{\varphi(N)}\lim_{R \rightarrow 0}  R^2 \int_{\mathcal{F}_{N}}\frac{d^2 \tau}{\tau_{2}^{2}} f(\tau, \bar{\tau})\sum_{(k,N) = 1}^{N-1}  % \sum_{m=-\infty}^{\infty}\sum_{n=-\infty}^{\infty}  e^{-\frac{\pi R^2}{N^{2}\tau_2}| nN\tau + m N + k|^2} \nn \\
&=& \frac{2}{\varphi(N)}\lim_{R \rightarrow 0}  R^2 \int_{0}^{\infty} \frac{d \tau_{2}}{\tau_{2}^{2}} \int_{-1/2 }^{1/2 } d\tau_1 f(\tau, \bar{\tau})\sum_{(s,N)=1}^{N-1}\sum_{r=-\infty }^{\infty}e^{-\frac{\pi R^2 (Nr + s)^2}{N^{2}\tau_2}} \nn \\
&=& \frac{2}{L(S^{1}(\tau_2))\varphi(N)}\lim_{R \rightarrow 0}  R^2 \int_{0}^{\infty} d \tau_{2} \oint_{S^{1}(\tau_{2})}ds f(\tau, \bar{\tau}) \sum_{(s,N)=1}^{N-1}\sum_{r=-\infty }^{\infty}e^{-\frac{\pi R^2 (Nr + s)^2}{N^{2}\tau_2}},  \nn \\ \label{unfolded2}
\eea
where $S^{1}(\tau_2)$ is the circle $S^{1}(\tau_2) = \{ -1/2 \le x <1/2, \  y = \tau_2   \} \subset \mathbb{H}/T$.

Notice that the length   of $S^{1}(\tau_2)$, $L(S^{1}(\tau_2)) = 1/\tau_2$  becomes infinite as $\tau_{2} \rightarrow 0$ due to the hyperbolic metric $ds = \sqrt{d\tau_{1}^2 + d\tau_{2}^2 }/\tau_{2}$ of
$\mathbb{H}$.

\vspace{.2 cm}
  In the Laurent  expansion iii) for $f$, the simple poles  in the cusps $q=0$ and on some points of the circle $|q|=1$
 may spoil eq. (\ref{unfolded2}). This is best seen for a divergence at the cusp $q=0$, ($\tau = i\infty$).
 In fact,  the integral of $f$ on the  region (figure \ref{domain}) $\mathcal{F} \subset \mathcal{F}_N$ which extends to $\tau = \infty$
  is convergent only  with   the prescription to perform the $\tau_1$ integral first, which eliminates
   $1/q$.

    \vspace{.1 cm}

 Since  in the limit $\tau_2 \rightarrow \infty$ the  exponential  factor in  (\ref{a})
behaves as
\beq
e^{-\frac{\pi R^2}{N^{2}\tau_2}|nN\tau +  m N + k|^2} \sim e^{-\pi R^2  n^{2} \tau_2}, \label{factor2}
\eeq
 the integral in  the first line of  the following equation

 \bea
 \int_{\mathcal{F}_N}\frac{d^{2}\tau}{\tau_{2}^2}f(\tau, \bar{\tau}) &=&
 \frac{1}{\varphi(N)}\lim_{R \rightarrow 0}R^{2} \int_{\mathcal{F}_{N}}\frac{d^{2}\tau}{\tau_{2}^2}f(\tau,\bar{\tau})\sum_{m=-\infty}^{\infty}\sum_{n=-\infty}^{\infty}\sum_{(k,N)=1}^{N-1}e^{-\frac{\pi R^{2}}{N^{2}\tau_2}|nN\tau + m N +k|^2} \nn \\
 &=&
\frac{2}{\varphi(N)}\lim_{R \rightarrow 0}  R^2 \int_{0}^{\infty} d \tau_{2} \int_{-1/2}^{1/2}d\tau_{1} f(\tau, \bar{\tau}) \sum_{(s,N)=1}^{N-1}\sum_{r=-\infty }^{\infty}e^{-\frac{\pi R^2 (Nr + s)^2}{N^{2}\tau_2}} \label{an}
 \eea
 is actually
absolutely convergent for $\tau_2 \rightarrow \infty$ for large enough $R > R_0$,  (the asymptotic  factor (\ref{factor2}) for  $R$
 large enough  cancels  the exponential growing factor
 $1/|q| = e^{2\pi\tau_2} $
 in the Fourier expansion of $f(\tau ,\bar{\tau})$ allowed by  condition iii)).

\vspace{.2 cm}

However, in order to take the limit for $R \rightarrow 0$, one needs equation (\ref{an})
to hold until $R>0$ and not just for $R > R_0$.
The validity of eq. (\ref{an}) on the full semi-axis $R>0$
can be checked  by considering  eq. (\ref{an})
for complex  $R$.

By using Poisson resummation formula on the first   line of
of (\ref{an}), one can rewrite this equation   in the following equivalent way

\bea
 \int_{\mathcal{F}_{N}}\frac{d^2 \tau}{\tau_{2}^{2}}f(\tau,\bar{\tau})
 &=& \frac{1}{\varphi(N)}\lim_{R \rightarrow 0} R \int_{\mathcal{F}_{N}}\frac{d^2 \tau}{\tau_{2}^{3/2}} f(\tau, \bar{\tau})\sum_{(k,N)=1}^{p-1} \sum_{m,n}e^{2\pi i k m/N}e^{2\pi imn\tau_1}
 e^{-\pi \tau_2\left( \frac{m^2}{R^2} + n^{2}R^{2} \right)}  \nn \\
&=& \frac{2}{\varphi(N)} \lim_{R \rightarrow 0} R^2 \int_{0}^{\infty} d \frac{\tau_{2}}{\tau_{2}^2}  \int_{-1/2}^{1/2} d\tau_{1}  f(\tau, \bar{\tau})\sum_{(s,r)=1}^{N-1}\sum_{r=-\infty}^{\infty}e^{-\frac{\pi R^2 (Nr+ s)^2}{N^{2}\tau_2}}. \nn \\ \label{unfolded3}
\eea

 The function in the  first line of (\ref{unfolded3}) is analytic in the complex variable  $R$ on a region
 where   the integral  converges as well as all its $R$-derivatives.
A breakdown of analyticity in $R$ happens whenever in   the Fourier expansion of the full integrand function
there is a point $R = \bar{R}$ where  a term non-exponentially suppressed for $\tau_2 \rightarrow \infty$
 appears.  By taking enough $R$-derivatives
 one would find a divergence in the integral for $\tau_2 \rightarrow \infty$   in such a point
 \footnote{This situation is formally equivalent to the lack of analyticity
 for the free-energy in a compactification that happens whenever
 for a certain value of a modulus a massless state appears. In that case
 this is a signal of a a possible phase transition, in the present
 case a lack of analyticity in $R$ may invalidate eq. (\ref{an}) for small $R$.}.

Since the factor at the exponent  in the  first line of (\ref{unfolded3}) for both $m$ and $n$ non-zero
satisfies
\beq
\frac{m^2}{R^2} + n^{2}R^{2} \ge 2|m n| \ge 2,  \label{bound}
\eeq

indeed terms proportional to $1/q$  in the Laurent expansion
for $f$  do spoil analiticity in the point $R = 1$, and invalidate (\ref{an}) for $0 <R \le 1$.

\vspace{.4 cm}

In order to avoid this problem we regularize the $f(\tau,\bar{\tau})$ at the cusps in a $\Gamma_{0}[N]$
invariant way. For example at the cusp $\tau = \infty$ we regularize $f \rightarrow \tilde{f}_{\infty}$ as follows
\beq
\tilde{f}_{\infty}(q,\bar{q}) = f(q,\bar{q}) - c_{\infty}J(q), \label{reg}
\eeq
where $J(q)$ is the Klein modular invariant function with Laurent expansion
\beq
J(q) =  \frac{1}{q}  + 196884q + 21493760q^2 +... =     \frac{1}{q} + \sum_{n=1}^{\infty}a_{n}q^{n}.
\eeq

For  a simple pole at a cusp $\tau_i \in \mathbb{Q} \cap [-1/2,1/2]$, $f$ is regularized for $q \rightarrow e^{2\pi i \tau_{i}}$
by
\beq
\tilde{f}_{\tau_{i}}(q,\bar{q}) = f(q,\bar{q}) - \beta_{i}c_{i}J(q).
\eeq

Since $J(q)$  has a simple pole at the cusp $\tau = \infty$, by modular invariance
it has simple poles in all the images of $\tau = \infty$ through modular transformations. In particular
$J(q)$ has simple poles  in all  the rational points in $[-1/2,1/2]$.

Moreover, $J(q)$ being holomorphic in $q$, it gives zero when integrated in $\tau_1$ on the interval $[-1/2,1/2]$.
Therefore it doesn't contribute to the integral along the one-dimensional curve,
while its contribution over a fundamental domain of $\mathcal{F}$ is

\beq
 \int_{\mathcal{F}} \frac{d^2\tau}{\tau_{2}^2} J(\tau) = -24\frac{\pi}{3}.
\eeq

Moreover, since

\beq
\int_{-1/2}^{1/2}d\tau_1 J(\tau) = 0
\eeq
for every $\tau_2$, the $J$ integral over $\mathcal{F}$ receives\footnote{The value $-24\pi/3$ of the integral of $J$ over the region $\mathcal{F}$ was computed in \cite{Moore:1987ue},\cite{Lerche:1987qk}.}   contribution only from the region $\aleph = \mathcal{F} - [-1/2, 1/2]\times [1, \infty)$
\beq
 \int_{\mathcal{F}} \frac{d^2\tau}{\tau_{2}^2} J(\tau) =
 \int_{\aleph} \frac{d^2\tau}{\tau_{2}^2} J(\tau) = -24\frac{\pi}{3}.
\eeq

Interesting enough, from the string theory point of view  $\aleph$ is the subregion of $\mathcal{F}$ where level matching is not enforced.
This is a peculiar characteristic  of string theory, since in field theory the proper time integration domain would have a rectangular shape.

\vspace{.4 cm}

Since eq. (\ref{unfolded2}) is valid for $\tilde{f}$ up to $R = 0$, one
 change of integration variable $\tau_2 \rightarrow R^{2}\tau_2$ and finally compute the $R \rightarrow 0$ limit
\bea
 \int_{\mathcal{F}_{N}}\frac{d^{2}\tau}{\tau_{2}^{2}}\tilde{f}(\tau,\bar{\tau}) &=& \frac{1}{\varphi(N)}\lim_{R \rightarrow 0}   \int_{0}^{\infty}  \frac{d\tau_{2}}{\tau_{2}^2} \oint_{S^{1}(R^{2}\tau_{2})}\frac{ds}{L(S^{1}(R^{2}\tau_{2}))} \tilde{f}(\tau, \bar{\tau}) \sum_{(s,N)=1}^{N-1}\sum_{r=-\infty }^{\infty}e^{-\frac{\pi  (Nr + s)^2}{N^{2}\tau_2}} \nn \\
&=& \frac{1}{\varphi(N)}\lim_{R \rightarrow 0}   \oint_{S^{1}(R^{2})}\frac{ds}{L(S^{1}(R^{2}))}      \tilde{f}(\tau, \bar{\tau})         \int_{0}^{\infty} d x \sum_{(s,N)=1}^{N-1}\sum_{r=-\infty }^{\infty}e^{-\frac{\pi  (Nr + s)^2 x}{N^{2}}} \nn \\
&=& \frac{2N^2}{\pi \varphi(N)}\left( \sum_{(s,N)=1}^{N-1}\sum_{n= 0}^{\infty} \frac{1}{(Nn + s)^2}\right) \lim_{R \rightarrow 0} \frac{1}{L(S^{1}(R^{2}))}  \oint_{S^{1}(R^{2})}ds \tilde{f}(\tau, \bar{\tau}) \nn \\
&=& \frac{2N^2}{\pi \varphi(N)} \left( \sum_{(s,N)=1}^{N-1}\sum_{n= 0}^{\infty} \frac{1}{(Nn + s)^2}\right)  \lim_{\tau_2 \rightarrow 0}  \int_{-1/2}^{1/2}d\tau_1 \tilde{f}(\tau, \bar{\tau}). \nn \\ \label{final}
\eea

By using
\beq
\int_{\mathcal{F}_N}\frac{d^{2}\tau}{\tau_{2}^2}\tilde{f}(\tau,\bar{\tau}) =
 \int_{\mathcal{F}_N}\frac{d^{2}\tau}{\tau_{2}^2}f(\tau,\bar{\tau} ) -  24\frac{\pi}{3}\sum_{i=1}^{n_{c}(N)}c_{i}\beta_{i},
\eeq
and
\beq
\int_{-1/2}^{1/2}d\tau_{1}J(\tau) = 0,
\eeq
one finally recovers
\beq
\int_{\mathcal{F}_{N}}\frac{d^2 \tau}{\tau_{2}^{2}}f(\tau,\bar{\tau}) = \frac{2N^2}{\pi \varphi(N)} \left( \sum_{(s,N)=1}^{N-1}\sum_{n= 0}^{\infty} \frac{1}{(Nr + s)^2}\right)  \lim_{\tau_2 \rightarrow 0}  \int_{-1/2}^{1/2}d\tau_1 f(\tau, \bar{\tau}) + 24\frac{\pi}{3}\sum_{i=1}^{n_{c}(N)}c_{i}\beta_{i},
\eeq
which proves the theorem.

\vspace{.4 cm}

The numerical factor in front of the  limit in (\ref{final}),
when $N$ is prime $N = p$ is given by

\bea
&&\frac{2p^{2}}{\pi \varphi(p)}\sum_{(s,p)=1}^{p-1} \sum _{n = 0}^{\infty }\frac{1}{ (Nn+s)^2} =
\frac{2p^{2}}{\pi (p-1)}\sum_{s=1}^{p-1} \sum _{n = 0}^{\infty }\frac{1}{ (Nn+s)^2} \nn \\
&=& \frac{2p^{2}}{\pi (p-1)}\sum _{n = 1}^{\infty }\left( \frac{1}{n^2} - \frac{1}{(p n)^2}  \right) \nn \\
&=& = \frac{2p^{2}}{\pi (p-1)}\left(1- \frac{1}{p^2}   \right)\sum _{n = 1}^{\infty } \frac{1}{n^2} = (p+1)\frac{\pi}{3}. \nn \\
\eea

$(p+1)\pi/3$ is the invariant area of the region $\mathcal{F}_{p}$ since $\mathcal{F}_{p} = \mathcal{F} \cup \bigcup_{i=0}^{p-1}ST^{i}(\mathcal{F})$,
for prime $p$.
Therefore one expects  for  generic $N$ the following  series  to   compute the area of the fundamental region of $\mathcal{F}_{N}$

\beq
A(\mathcal{F}_{N}) = \frac{2N^{2}}{\pi \varphi(N)}\sum_{(s,N)=1}^{N-1} \sum _{n = 0}^{\infty }\frac{1}{ (Nn+s)^2}. \label{FN}
\eeq

  For  every  positive integer $N$, $\Gamma_{0}[N] \subset \Gamma$, and therefore $ \mathcal{F} = \mathbb{H}/\Gamma \subset     \mathcal{F}_{N} = \mathbb{H}/\Gamma_{0}[N]$.
 Indeed, the fundamental region  $\mathcal{F}_{N}$ is  tassellated by a finite number $\mathcal{N}(N)$ of
 fundamental regions of  $\Gamma$, each region with invariant area $\pi/3$.
 
 Therefore from eq. (\ref{FN}) one obtains the number of  $\mathcal{F}$-tiles $\mathcal{N}(N) \in \mathbb{N}$
 needed to cover $\mathcal{F}_{N}$
 \beq
\mathcal{N}(N) = \frac{A(\mathcal{F}_{N})}{A(\mathcal{F})}= \frac{6N^{2}}{\pi^{2} \varphi(N)}\sum_{(s,N)=1}^{N-1} \sum _{n = 0}^{\infty }\frac{1}{ (Nn+s)^2}.
\eeq

 The sequence $\{\mathcal{N}(N) \}_{N}$ starts with
\beq
\{ 1,3,4,6,6,12,8,12,12,18,12,24,14,24,24,18,36,20,36,...\}. \nn
\eeq

$\mathcal{N}(N)$  drops down in correspondence of  prime numbers,  $\mathcal{N}(p) \le  \mathcal{N}(p-1)$ for $p$ prime. In fact the congruence subgroups for prime numbers
are larger then the non-prime adjacent ones, and this difference  becomes more relevant for large $N$.


\begin{thebibliography}{99}


%%%%%%%%%%%%%%%%%%%%%%%%%%%%
%Moduli Stabilization
%%%%%%%%%%%%%%%%%%%%%%%%%%%%%%


%%%%%%%%%%%%%%%%%%%%%%%%%%%%%%%%%%%%%
%Torus stabilization
%%%%%%%%%%%%%%%%%%%%%%%%%%%%%%%%%%%%
\bibitem{Angelantonj:2006ut}
  C.~Angelantonj, M.~Cardella and N.~Irges,
  ``An Alternative for Moduli Stabilisation,''
  Phys.\ Lett.\  B {\bf 641}, 474 (2006)
  [arXiv:hep-th/0608022].



\bibitem{Dine:2006gx}
  M.~Dine, A.~Morisse, A.~Shomer and Z.~Sun,
  ``IIA moduli stabilization with badly broken supersymmetry,''
  JHEP {\bf 0807}, 070 (2008)
  [arXiv:hep-th/0612189].


\bibitem{Angelantonj:2008fz}
  C.~Angelantonj, C.~Kounnas, H.~Partouche and N.~Toumbas,
  ``Resolution of Hagedorn singularity in superstrings with gravito-magnetic
  fluxes,''
  Nucl.\ Phys.\  B {\bf 809}, 291 (2009)
  [arXiv:0808.1357 [hep-th]].






%%%%%%%%%%%%%%%%%%%%%%%%%%%%%%%%%%%%%%%%unfolding%%%%%%%%%%%%%%%%%%%%%%%%%%%%%%%%%%%%%%%%%%%%%%%%%%%%%%%%%%%%%%%

\bibitem{McClain}  B. McClain and B. D. B. Roth, Modular Invariance For Interacting Bosonic Strings At Finite
Temperature, Commun. Math. Phys. 111 (1987) 539.


\bibitem{fBrien} K. H. OfBrien and C. I. Tan, Modular Invariance Of Thermopartition Function And Global
Phase Structure Of Heterotic String, Phys. Rev. D36 (1987) 1184.

\bibitem{Itoyama} H. Itoyama and T. R. Taylor, Supersymmetry Restoration In The Compactified O(16)~OŒ(16)
Heterotic String Theory, Phys. Lett. B186 (1987) 129.

\bibitem{Dixon} L. J. Dixon, V. Kaplunovsky and J. Louis, Moduli dependence of string loop corrections to
gauge coupling constants, Nucl. Phys. B355 (1991) 649.

\bibitem{MS} P. Mayr and S. Stieberger, Threshold corrections to gauge couplings in orbifold compactifications,
Nucl. Phys. B407 (1993) 725 [arXiv:hep-th/9303017].

\bibitem{GNS} D. M. Ghilencea, H. P. Nilles and S. Stieberger, Divergences in Kaluza-Klein models and
their string regularization, New J. Phys. 4 (2002) 15 [arXiv:hep-th/0108183].

\bibitem{Tra} M. Trapletti, On the unfolding of the fundamental region in integrals of modular invariant
amplitudes, JHEP 0302 (2003) 012 [arXiv:hep-th/0211281].

\bibitem{KKPR} E. Kiritsis, C. Kounnas, P. M. Petropoulos and J. Rizos, String threshold corrections in
models with spontaneously broken supersymmetry, Nucl. Phys. B540 (1999) 87 [arXiv:hepth/
9807067].

%%%%%%%%%%%%%%%%%%%%%%%%%%%%%%%%%%%%%%%%%%%%%%%%MATH%%%%%%%%%%%%%%%%%%%%%%%%%%%%%%%%%%%%%%%%%%%

\bibitem{F} H. Furstenberg, The Unique Ergodicity of the Horocycle Flow, Recent Advances in Topological Dynamics, A. Beck (ed.), Springer Verlag Lecture Notes, 318 (1972), 95-115.

\bibitem{Dani} S. G. Dani and J. Smillie, Uniform distribution of horocycle orbits for Fuchsian groups. Duke Math. J. 51 (1984), 185–194.

\bibitem{Ratner1} M. Ratner,   Distribution rigidity for unipotent actions on homogeneous spaces.  Bull. Amer. Math. Soc. (N.S.) Volume 24, Number 2 (1991), 321-325.

\bibitem{Ratner2} M. Ratner,  Raghunathan's topological conjecture and distributions of unipotent flows.  Duke Math. J. Volume 63, Number 1 (1991), 235-280.





%%%%%%%%%%%%%%%%%%%%%%%%%%%%%%%%%%%%%%%%%%%%%%%%%%%%%%%%%%%%%%%%%%%%%%%%%%%%%%%%%%%%%%%%%%%%%%%%%%%%%%%%%%%%%
%Asymptotic SUSY
%%%%%%%%%%%%%%%%%%%%%%%%%%%%%%%%%%%%%%%

%\cite{Kutasov:1990sv}
\bibitem{Kutasov:1990sv}
  D.~Kutasov and N.~Seiberg,
``Number Of Degrees Of Freedom, Density Of States And Tachyons In String
  Theory And Cft,''
  Nucl.\ Phys.\  B {\bf 358}, 600 (1991).

%\cite{Kutasov:1991pv}
\bibitem{Kutasov:1991pv}
  D.~Kutasov,
  ``Some properties of (non)critical strings,''
  arXiv:hep-th/9110041.
  %%CITATION = HEP-TH/9110041;%%

%\cite{Dienes:1995pm}
\bibitem{Dienes:1995pm}
  K.~R.~Dienes, M.~Moshe and R.~C.~Myers,
  ``String Theory, Misaligned Supersymmetry, And The Supertrace Constraints,''
  Phys.\ Rev.\ Lett.\  {\bf 74}, 4767 (1995)
  [arXiv:hep-th/9503055].
  %%CITATION = PRLTA,74,4767;%%

%\cite{Dienes:1994es}
\bibitem{Dienes:1994es}
  K.~R.~Dienes,
  ``Modular invariance, finiteness, and misaligned supersymmetry: New
  constraints on the numbers of physical string states,''
  Nucl.\ Phys.\  B {\bf 429}, 533 (1994)
  [arXiv:hep-th/9402006].
  %%CITATION = NUPHA,B429,533;%%

%%%%%%%%%%%%%%%%%%%%%%%%%%%%%%%%%%%%%%%%%%%%%%%
%J
%%%%%%%%%%%%%%%%%%%%%%%%%%%%%%%%%%%%%%%%%%%%%
%\cite{Moore:1987ue}
\bibitem{Moore:1987ue}
  G.~W.~Moore,
  ``Atkin-Lehner Symmetry,''
  Nucl.\ Phys.\  B {\bf 293} (1987) 139
  [Erratum-ibid.\  B {\bf 299} (1988) 847].
  %%CITATION = NUPHA,B293,139;%%

%\cite{Lerche:1987qk}
\bibitem{Lerche:1987qk}
  W.~Lerche, B.~E.~W.~Nilsson, A.~N.~Schellekens and N.~P.~Warner,
  %``ANOMALY CANCELLING TERMS FROM THE ELLIPTIC GENUS,''
  Nucl.\ Phys.\  B {\bf 299}, 91 (1988).
  %%CITATION = NUPHA,B299,91;%%


\end{thebibliography}
\end{document}